\documentclass[a4paper,11pt]{article}
\usepackage{amssymb}  
\usepackage{graphicx}
\usepackage{geometry}
\providecommand{\opusversion}{version 1.51, \today}

\begin{document}

\title{Star -- triangle equivalence in soap froths}

\author{
M. Mancini,
C. Oguey\thanks{Email: oguey@ptm.u-cergy.fr}\\
LPTM\thanks{CNRS UMR 8089}, Universit\'e de Cergy-Pontoise, 95031 Cergy-Pontoise, France
}
\date{\opusversion} 

\maketitle

\begin{abstract}
In two dimensional foams at equilibrium, triangular bubbles can be freely exchanged with 3-fold stars ---three edges ending at a central vertex.
This theorem is deduced here from Moukarzel's duality.
Moreover, to probe the method, a few related properties are established: under slow gas diffusion, T2 processes are continuous for triangles but not for other types of bubbles. In general, the gas flow results in different configurations in the presence of a triangle than in the presence of a star.
\end{abstract}


\section{Foam equilibrium}

In a two dimensional foam at equilibrium, the films are arcs of circles meeting 3 by 3 at vertices. The first property is Laplace law with uniform gas pressure inside each bubble. The second is stability or genericity, viewing foams as random patterns \cite{weaireRiv,weaireHut}.
Two further conditions characterize equilibrated foams:
\begin{itemize}
\item The sum of the edge curvatures $k_{ij}$ around any vertex is zero (Laplace).
\item At any vertex, the angles between the films are $\frac{2\pi}{3}$ (Plateau).
\end{itemize}
These Plateau-Laplace conditions follow from mechanical equilibrium or energy minimisation
under the constraints of constant amount of matter in each bubble. In the incompressibility approximation, the energy of foams is proportional to the total length of the films, or perimeter.

More general situations can be considered where the surface tension has different values $\gamma_{ij}$ on different sides. Then the equilibrium P-L conditions become
\begin{enumerate}
\item\label{laplaceG} $\sum \gamma_{ij}\, k_{ij}=0$ around any vertex.\hfill(L)
\item\label{plateauG} $\sum \gamma_{ij}\, \vec{t}_{ij}=0$ at any vertex,\hfill(P)
\end{enumerate}
where $\vec{t}_{ij}$ are unit vectors tangent to the films at the considered vertex.

As described so far, the model seems restricted to foams in the dry limit. However, this is not the case. The model is also valid for foams with a small liquid fraction. Indeed, the liquid is confined into concave triangular regions around the vertices ---the Plateau borders--- and the film circular arcs may be prolongated inside the Plateau borders so as to meet three by three in a way satisfying Plateau's rule. This {\em vertex decoration lemma} \cite{boltonWeaire} or {\em theorem} \cite{weaire} follows from hydrostatic equilibrium. The restriction comes only from the edges that are too short to support separated liquid triangles at their ends; this gives rise to borders with more than 3 (curved) sides and the simplified geometry fails.

We call it ``vertex'' decoration to distinguish it from a {\em bubble decoration theorem} which is our subject here. It may be summarised as follows: any three-sided bubble can be considered as a decoration of a virtual vertex which is the unique intersection of the three external edges incident to the triangle, suitably prolonged. Moreover, Plateau's rule is satisfied at this vertex. In other words, any triangular bubble my be replaced by a star: a vertex with the circular arcs needed to complete the 3 external edges; reciprocally, any vertex may be decorated by substituting to it a triangular bubble, small enough not to enclose any entire edge from the original three star. The equilibrium is preserved by any of these star-triangle transformations.

In more details, this property contains two parts: 1) when prolonged into the triangle, the three external edges meet at a common point, the star point or vertex; 2) the angles of the prolonged edges at the star point satisfy Plateau.

The main corollary following this theorem is continuity at T2 processes. Suppose that the foam evolves slowly enough that assuming equilibrium at each time is a good approximation. Evolution is due to gas diffusion through the weakly permeable films, eventually with quasistatic external constraints.
When a side becomes vanishingly short, two vertices meet and, when they coalesce, the configuration may not be a foam equilibrium any more. Therefore a fast evolution occurs, energy dissipates until a new quasi-equilibrium is reached. 
It is customary to decompose such topological changes into elementary processes: edge switch T1, preserving the number of cells, and cell extinction T2.

When a 3-sided bubble shrinks, the outside edges extend to a (virtual) equilibrated star figure at each time. So, precisely when the triangle degenerates to a point, the star vertex becomes real and the figure is {\it ipso facto} in equilibrium. No fast evolution, or discontinuity at the slow scale, is needed.
Compare this with the discontinuity generally implied in T1 \cite{boltonWeaireII} processes or in T2 ones when the extinguishing bubble has 4 sides or more \cite{fradkovMUW}. 

\section{A bubble decoration theorem}

We now show that, under the hypothesis of P-L equilibrium, the three external edges ---legs, for short--- of a three-bubble $C_1$ extend to a common point and that they meet in way satisfying Plateau's rule.

We give two proofs, one using simple geometry, the other based on Moukarzel's reciprocity.

The first proof uses reflection, valid when the angles all equal $\frac{2\pi}{3}$. So constant surface tension is assumed.
\begin{figure}[ht]
\centering 
\begin{tabular}{ccc}
\includegraphics[height=6cm]{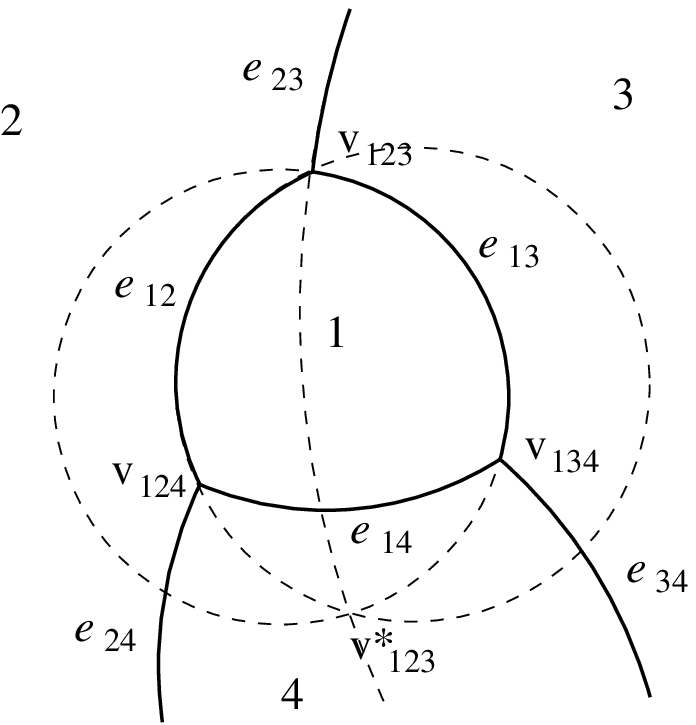}  &\quad& 
\includegraphics[height=5cm]{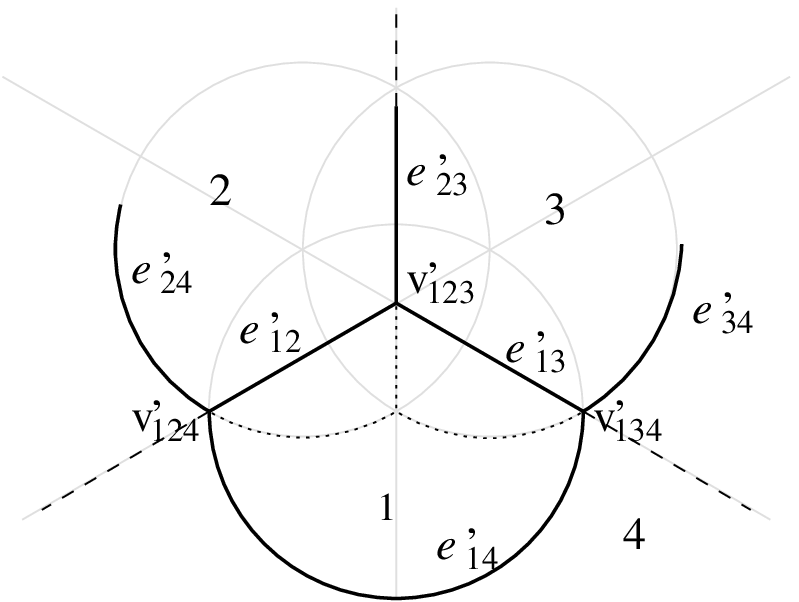} \\  
a) & & b)
\end{tabular}
\caption{a) A generic three-edge bubble contained in an equilibrated foam.\label{triangle1}
b) The same bubble transformed by a suitable inversion.\label{triangle2} }
\end{figure}

\subsection{Inversion}
The symmetry group of equilibrated foams is $SL(2,\mathbb{C})$, the group of conformal transformations of the compactified plane $\overline{\mathbb{C}}=\mathbb{C} \cup \{\infty \}$, conformally equivalent to the sphere $S^2$. This is an extension of the 2D Euclidean group by inversions\,\footnote{Also called homographies, Moebius or linear fractional transformations.}. Beside preserving angles (conformality), these transformations map circles to circles; in particular any circle passing through the inversion point $v^*$ is sent to a straight line (circle through infinity). Moreover the equilibrium conditions (P-L) are conserved by these transformations \cite{moukarzel,weaire}.

\subsection{Local reflection principle}
A straight edge is a symmetry axis for the figure formed by any of its end vertices and the two circles incident to that vertex (the edge itself is also symmetric, of course).
Indeed, the circles are defined by the vertex $v$, the tangent at that point and curvature. The tangents are symmetric by Plateau; equality of curvatures (up to sign) follows from Laplace and the value zero of the curvature of the straight edge.
In particular the line defined by the centres of the two circles through $v$ is orthogonal to the straight edge.
This reflection principle is only valid when the surface tensions have the same value.

\subsection{Proof 1}

A crucial property of equilibrated foams is alignment \cite{moukarzel,weaire}: the centres of the 3 (arcs of) circles meeting at any vertex $v$ are aligned. This is equivalent to the existence of a point  $v^*$, conjugate to $v$, where the three circles also meet. $v^*$ is symmetric to $v$ with respect to the line of the centres.
When one of the edges is straight, alignment reduces to orthogonality, as implied by local reflection symmetry.

Let us apply an inversion transformation with respect to the point $v^*_{123}$, conjugate of $v_{123}$. In the settings of figure~\ref{triangle1}, the circles of $e_{23}$, $e_{13}$ and $e_{12}$ are mapped onto lines, $e'_{23}$, $e'_{13}$ and $e'_{12}$ meeting at $v'_{123}$ with angles of $\frac{2\pi}{3}$ (figure~\ref{triangle2}b).

We can assume that $v'_{123}$ is at the origin $0$.
Consider the vertex $v{\,}_{\!124}'$ on the line $e'_{12}$. As the angles of $\frac{2\pi}{3}$ are preserved, the tangent to $e'_{14}$ at $v'_{124}$ must be vertical (parallel to $e'_{23}$). So the centre $c'_{14}$ of $e'_{14}$ is on the same horizontal line as $v'_{124}$.
By the same argument at $v'_{134}$, we find that $v'_{134}$, $v'_{124}$ and $c'_{14}$  are on the same horizontal line. Therefore $e'_{14}$ is a half circle centred on the axis $e'_{23}$ which, thus, becomes a symmetry axis for the triangle. The reflection principle can now be applied to the two edges $e'_{12}$ and $e'_{13}$ to extend the symmetry to the figure made of the triangle and the circles supporting its legs. This figure becomes completely symmetric under the group $D_3$ (generated by mirrors at $\pi/3$).

In particular, the two circles supporting $e'_{24}$ and $e'_{34}$ intersect at a point on the axis $e'_{23}$ ---proving convergence--- and they meet with angle $\frac{2\pi}{3}$ ---proving Plateau.
Applying the inverse transformation ensures the required properties for the original three-bubble (foam).

As a corollary, equilibrated triangles are rigid modulo homographies.
For example, by $SL(2,\mathbb{C})$, any equilibrated three-sided bubble can be mapped onto the regular equilateral curved triangle with angles $2\pi/3$ and straight external legs.

\subsection{Duality}
The 3-bubble decoration property is also true for general foams with edge dependent superficial tensions. Our second proof will be valid in this case.

It is based on the following theorem by Moukarzel \cite{moukarzel}:
a cellular pattern bounded by circle arcs meeting at threefold vertices represents an equilibrated foam $\mathcal{F}$ if and only if there is an oriented dual figure $\mathcal{F}^*$. Each bond in the dual is straight and must be orthogonal to its corresponding curved interface in the foam.

Under these conditions, the nodes of the dual may be considered as sources generating the foam by distance (in)equalities. In addition to its location, each source $P_j$ gets two parameters: an additive parameter, or altitude, controlling the size of the cell relative to its neighbours, and a multiplicative one controlling the curvatures. The cells are then the territories satisfying $d_i(x)\leq d_j(x)$ for all $j\neq i$, with an appropriate distance $d_j(\cdot)$ defined in terms of the Euclidean distance to the source  $P_j$ and its parameters. The precise form of the function $d_j$ ensures that the boundaries are circular arcs.

\subsection{Proof 2}
\begin{figure}[ht]
\centering
\includegraphics[width=14cm]{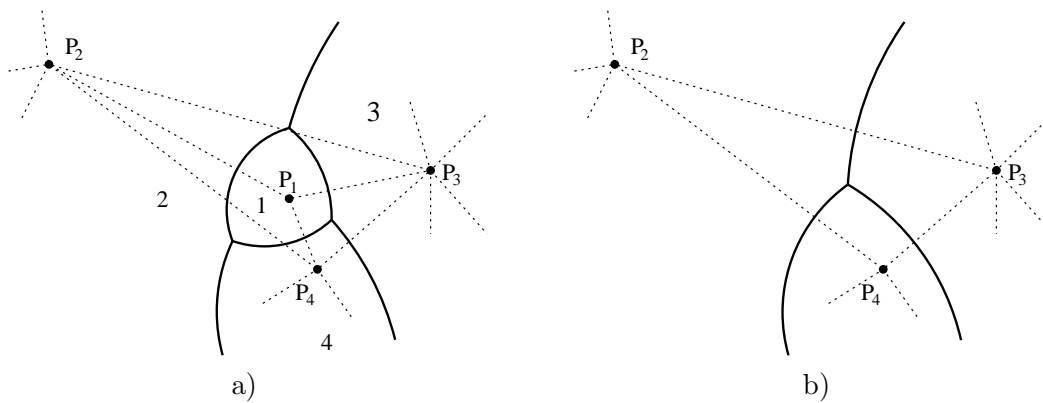}\\
a) \hspace{7cm} b)
\caption{ a) A three-edge bubble with its neighbours and the dual figure (dotted lines).
b) Same part without the three-edge bubble.} \label{star-tri}
\end{figure}

The 3 neighbours $C_2,\,C_3,\,C_4$ of a 3-bubble correspond to 3 sources $P_2,\, P_3,\, P_4$ forming a triangle in $\mathcal{F}^*$. The central bubble $C_1$ is represented by a point $P_1$ connected to the vertices of the triangle by three bonds (figure~\ref{star-tri}-a). Removing or inserting the three-bubble in $\mathcal{F}$ is the same as removing or inserting the source $P_1$ in $\mathcal{F}^*$ (figure~\ref{star-tri}-b).

Now the outer edges each satisfy an equality $d_i(x)= d_j(x)$ for a pair $i<j$ taken in $\{2,3,4\}$. When two edges meet at some $x_0$, say $d_2(x_0)= d_3(x_0)$ and $d_3(x_0)= d_4(x_0)$, then the third equality is also satisfied by transitivity. The condition on the angles follows from the fact that the vertices built in such a way are automatically equilibrated (see \cite{moukarzel}).

\section{Generalisations and discussion}

Duality, whence the decoration theorems, apply to a wide class of cellular patterns whose equilibrium is governed solely by surface tension.
Regarding pure equilibrium configurations, and in absence of any other field such as gravity, the cells filled with gas or liquid can be treated similarly.
So, as far as geometry is concerned, there is no fundamental distinction between vertex (Plateau borders) and bubble decorations.
The difference is only quantitative; up to small corrections due to disjoining pressure, the surface tension for the films is twice the value for liquid-gas interfaces (Plateau border boundaries).

A further extension is as follows: any cell cluster surrounded by three bubbles is connected to the rest of the foam by three edges ---its 'legs' as we said previously. The same argument would show that the legs are arcs of circles meeting at a common, equilibrated (virtual) vertex. In other words, star-triangle extends to {\em star--tripod} equivalence. 

\subsection{What about multipods~?}

\begin{figure}[ht]
\centering
\includegraphics[width=7cm]{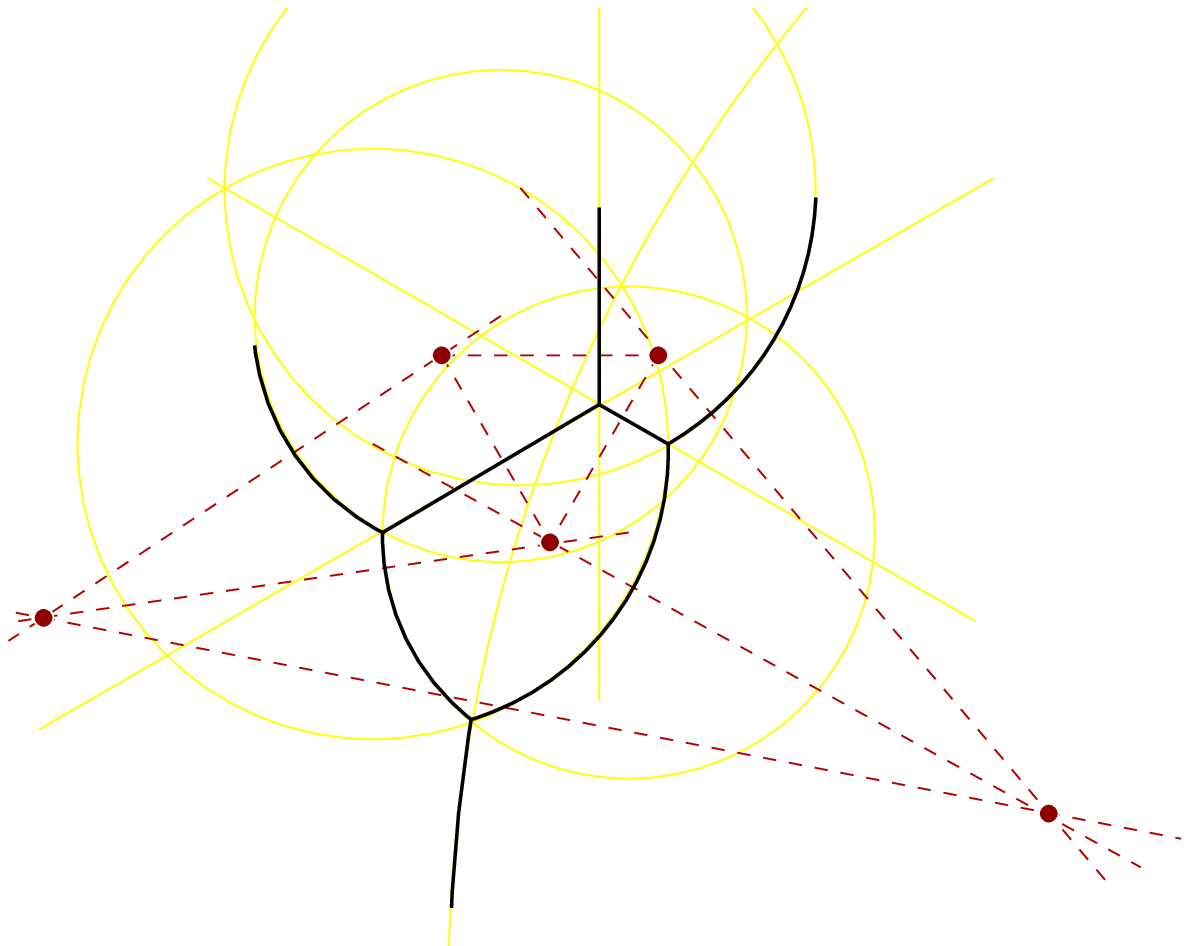}
\includegraphics[width=7cm]{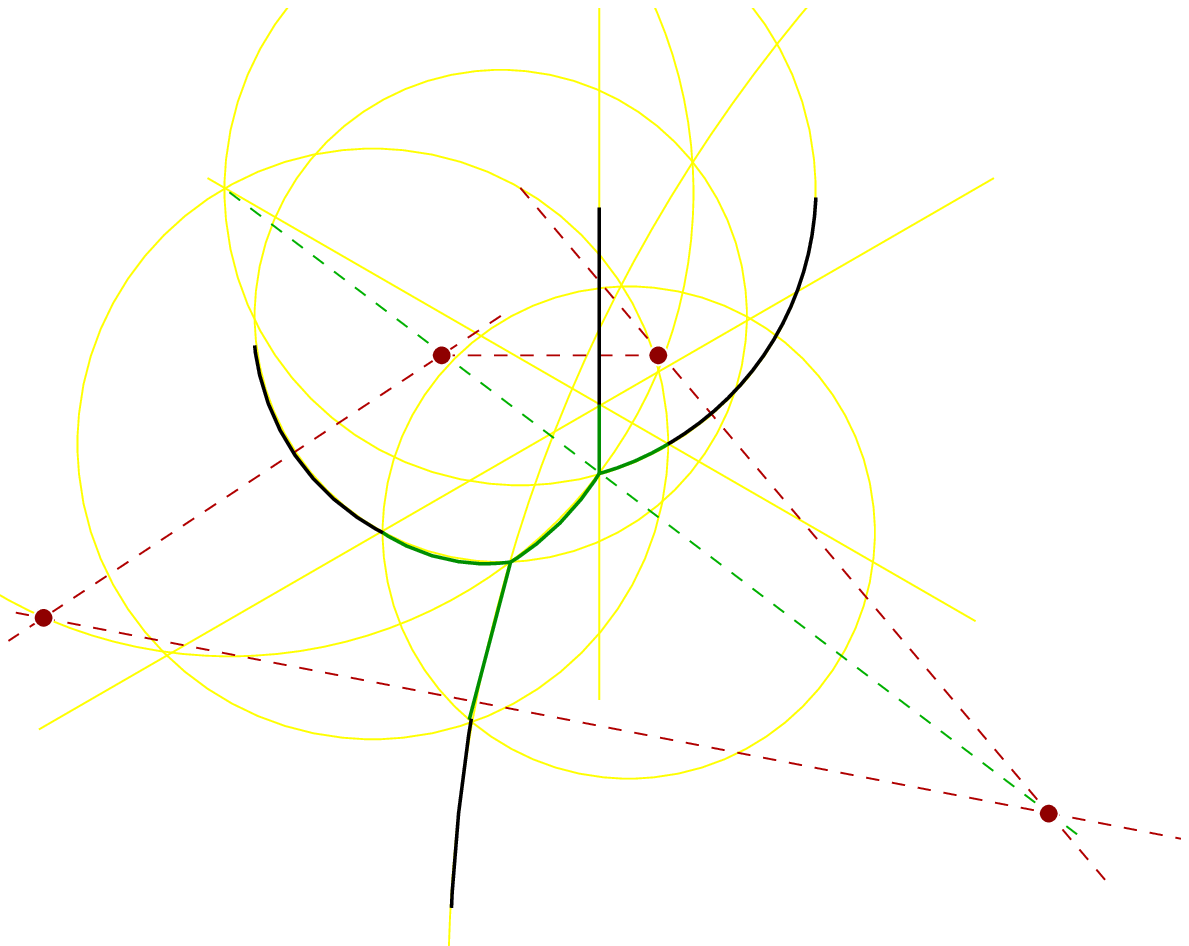}
\caption{A 4-bubble (left) transformed into a linked pair of vertices (right).
} \label{bub4}
\end{figure}

Can the same argument be applied to cells ---gas or liquid--- with more than 3 legs~?
Consider, for example, a 4 sided cell. In the reciprocal figure, it is represented by a source point connected to 4 neighbours. Removing this source without changing anything else will replace the 4-cell in the foam by a pattern made of two vertices joined by a curved edge, each of these vertices being connected to two of the former legs, suitably prolonged (figure~\ref{bub4}).
In the class of generalised model foams, which is the natural context of the theory, the new pattern is perfectly equilibrated. The trouble is that the value of the line tension, inherited by the new edge from the parameters of the dual sources, has nothing to do with any physically plausible value. Looking at figure~\ref{bub4}, the angles of the new edge with the legs are far form $2\pi/3$, apparently violating Plateau's law. More precisely, generalised Plateau (P) is still satisfied, but the theoretical value $\gamma_e$ required to make the new configuration a true equilibrium does not match the physical one, supposing one would realize the transformation physically. This mismatch renders the substitution of a 4-bubble by an H incompatible with equilibrium. When one of these bubbles shrinks, this mismatch persists down to zero area where it breaks equilibrium, triggers dynamical processes (fast or slow) and the foam needs to adjust its configuration in order to equilibrate the {\em physical} tensions.

\subsection{Coarsening}

\begin{figure}[ht]
\centering
\includegraphics[width=8cm]{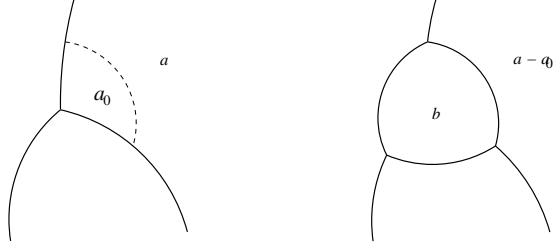}
\caption{A bubble of area $a$ gains one side and loses a portion of area $a_0$ in the presence of a neighbouring triangle. The area $b$ of the triangle is composed of 3 such portions.} \label{triangle3}
\end{figure}

The star-triangle equivalence is a pure equilibrium property that cannot be extrapolated to either mechanical response or time evolution. To demonstrate this point, let us compare two samples slowly evolving under gas diffusion, neglecting any effect from the Plateau borders. The two foams are identical at a time $t=0$ except for a star-triangle transformation. Let us also assume that no other topological changes occur during the life time $t_0$ of the triangular bubble.
Initially, if a bubble with $n$ sides and area $a$ has the star participating to its boundary, its replica neighbouring the triangle has $n+1$ sides and area $a-a_0$ (Figure \ref{triangle3}).
At any later time $t\leq t_0$, bubble area obeys Von Neumann's law \cite{vonNeumann}: $a(t)=a(0)-\kappa(6-n)t$ with $\kappa$ a constant. So, at the time $t_0=b/(3\kappa)$ when the triangle vanishes ---and the topology of the two samples converges---, the area of the neighbour is $a-\kappa(6-n)t_0$ in one case, $a-a_0-\kappa(6-n-1)t_0$ in the other. So there is a metrical difference of $\kappa t_0-a_0=b/3-a_0$ between the two situations: in general, replacing a triangle by a vertex changes the way the foam coarsens.

\subsection{Conclusion}

As presented here, star-tripod equivalence is one of the many properties that can be inferred, in a straightforward way, from Moukarzel's duality method \cite{moukarzel}.
This equivalence is related to the fact that all the triangular bubbles are equivalent under the symmetry group of foams, larger than the standard Euclidean group.
As a consequence, the vanishing of a triangle by diffusion is a geometrically continuous process.
Conversely, bubbles with $n\geq 4$ sides have intrinsic degrees of freedom which, in general, imply discontinuity, or breaking of equilibrium, at extinction (T2). 

As an application, star-triangle equivalence could be used to reduce the number of coordinates, thereby improving the efficiency of numerical minimisation, for example. As we have seen, this equivalence is only reliable for equilibrium questions. When the foam evolves, slowly or quickly, or when it is subjected to mechanical constraints, this equivalence is broken; triangles cannot be considered as decorations of vertices any more.

The natural extensions to dimension 3 concern cellular patterns whose boundaries are spherical patches. Albeit pleasant, this type of geometry is too restricted for real 3D foams. 

\section*{Acknowledgement}
We would like to thank N. Rivier for stimulating discussions.


\end{document}